\documentstyle[referee]{laa}

\begin{document}
\thesaurus{07  
     (07.09.1; 
      07.13.1;	
     )}

\title{THE POYNTING-ROBERTSON EFFECT AND SECULAR CHANGES OF ORBITAL
ELEMENTS}
\author{J.~Kla\v{c}ka}
\institute{Astronomical Institute,
    Comenius University, Mlynsk\'{a} dolina \\
    842~48 Bratislava, Slovak Republic;
    E-mail: klacka@fmph.uniba.sk}
\date{}
\maketitle

\begin{abstract}
The Poynting-Robertson (P-R) effect has been applied to meteoroids for several
decades. It is well known that the P-R effect produces only changes in the
orbital plane of the particle. The differential equations governing the secular
changes in both eccentricity and semi-major axis are known since the time of
Robertson, provided the initial orbits are not near circular. A disadvantage
of this type of osculating orbital elements is that it cannot be applied
to motion of arbitrarily shaped particles. Relevant type of osculating
orbital elements is discussed. It's application to the P-R effect is presented
and simple analytical formulae for secular changes in both eccentricity and
semi-major axis are derived.


\end{abstract}

\section{Introduction}
The relativistic equation of motion for a perfectly absorbing spherical dust
particle under the action of electromagnetic radiation was obtained by Robertson
(1937). Robertson derived expressions for the secular changes in the orbital
elements. The effect essentially reduces both semi-major axis and eccentricity
and is popularly known as the Poynting-Robertson (P-R) effect. The results of
Robertson were verified by Wyatt and Whipple (1950) and applied to the evolution
of meteoroid streams. One particularly important aspect, first pointed out by
Harwit (1963), is that a meteoroid escapes from the Solar System when the
orbital energy becomes positive and that this can happen when the energy due to
the radial component of the radiation force is included, without it being
necessary for this force to exceed the gravitational attraction.

A relativistic generalization corresponding to the P-R effect was discussed by
Kla\v{c}ka (1992a). A more simple derivation is given in Kla\v{c}ka (1993a).
There have been many misunderstanding, and even errors, in the treatment of
radiation force effects in the litterature and some of these are discussed in
Kla\v{c}ka (1993b, 2000b). We will not dwell further on most of these here.
A consistent set of equations governing, to the first order in $\vec{v} / c$,
the secular changes in the orbital elements were given by Kla\v{c}ka (1992b)
(here, $\vec{v}$ is velocity of the particle, $c$ is the speed of light),
where the appropriate initial conditions were also discussed. This work, and
most of the preceeding work assumes that the orbits are not near circular.
When the orbits are near circular, different considerations must apply, for
example a circular orbit can not reduce in radius without increasing in
eccentricity, contrary to the normal behaviour under P-R effect.
The near-circular situation has been discussed by Kla\v{c}ka and
Kaufmannov\'{a} (1992, 1993); the results analytically confirmed
Breiter and Jackson (1998) and their nonphysical conclusions are discussed in
Kla\v{c}ka (2000a, 2001a). We will not consider this near-circular case in
detail, but discussion will contain a short comment.

The Poynting-Robertson effect also causes an advancement of
perihelion, although the the effect is second order in $\vec{v} / c$
(that is, the secular change is proportional to $c^{-2}$).
Correct result can be found in Balek and Kla\v{c}ka (2002).

The aim of this paper is to discuss relevant form(s) of osculating orbital
elements and their secular changes. The motivation comes from the fact that the
P-R effect represents only a very special case of interaction between a particle
and incident electromagnetic radiation (Kla\v{c}ka 2000c, 2000d, 2000e, 2001b,
2002a, 2002c, Kla\v{c}ka and Kocifaj 2001). Type of secular changes of osculating
orbital elements derived by Robertson (1937) cannot be compared with secular
changes of these elements for real particles: secular changes of osculating
orbital elements based on central gravity force alone, are relevant physical
quantities (Kla\v{c}ka 2002b). The correct equations governing the secular
evolution of semi-major axis and eccentricity are derived for the P-R effect.

\section{P-R effect: Equation of motion}
Consider a spherical particle with a geometric cross section  $A'$ in
a flow of incident radiation with flux density $S$ in a direction
denoted by a unit vector $\vec{S}$.
Let $Q'_{pr}$ denote the efficiency factor for radiation. As
is usual within a relativistic context, we use $\gamma \equiv 1 /
\sqrt{1 - \vec{v}^{2} / c^{2}}$.
It is convenient to define $w = \gamma ( 1 - \vec{v} \cdot \vec{S} / c )$.

The relativistically covariant equation of motion for the P-R effect
may be expressed as (Kla\v{c}ka 1992a, 2000c, 2002a, 2002c)
\begin{equation}\label{1}
\frac{d ~p^{\mu}}{d~ \tau} = \frac{w^{2}~S~A'}{c^{2}} ~Q_{pr} ' ~ \left (
			     c ~ b^{\mu} ~-~ u^{\mu}  \right ) ~,
\end{equation}
where $p^{\mu}$ is four-vector of the particle of mass $m$,
$p^{\mu} = m~ u^{\mu}$,
four-vector of the world-velocity of the particle is
$u^{\mu} = ( \gamma ~c, \gamma ~ \vec{v} )$
and $b^{\mu} = ( 1 / w ) ~( 1, \vec{S} )$.

Relativistically covariant form of equation of motion is important in deriving
an expression for the secular change of an advancement of
perihelion, which can be found in Balek and Kla\v{c}ka (2002). Moreover,
higher orders in $v/c$ play an important role when making calculations for near
circular orbits (compare Breiter and Jackson 1998, Kla\v{c}ka 2000a, 2001a).
This will not be considered here, except section 3.1.
Thus, reduction of Eq. (1) to the first order in $\vec{v}/c$ will be used,
if it is not explicitly stressed.

Applying Eq. (1) to the situation where a particle is moving
in gravitational and electromagnetic fields
of a star (usually the Sun) with mass $M$ gives
\begin{equation}\label{2}
\frac{d\vec{v}}{dt} = - \frac{\mu}{r^{2}} \vec{S} + \beta \frac{\mu}{r^{2}}
    \left \{
    \left ( 1 - \frac{\vec{v} \cdot \vec{S}}{c} \right ) \vec{S}
    - \frac{\vec{v}}{c} \right \} ~,
\end{equation}
where $\mu = G M$, $G$ being the gravitational constant,
$\beta$ is the ratio of radiation pressure force to gravity force, that is
$\beta = Q'_{pr}~ A'~ r^{2} ~S~/~( c~ \mu~m )$.

\section{Secular changes of orbital elements -- radiation pressure
as a part of central acceleration}
We have to use $-~\mu ~( 1~-~\beta )~\vec{S}~/~r^{2}$ as a central
acceleration determining osculating orbital elements if we want to take a time
average ($T$ is time interval between passages through two following
pericenters) in an analytical way
\begin{eqnarray}\label{3}
\langle g \rangle &\equiv& \frac{1}{T}	\int_{0}^{T} g(t) dt =
\frac{\sqrt{\mu ~( 1 - \beta )}}{a_{\beta}^{3/2}} ~
 \frac{1}{2 \pi}  \int_{0}^{2 \pi} g(f_{\beta})
\left ( \frac{df_{\beta}}{dt} \right )^{-1} df_{\beta}
\nonumber \\
&=& \frac{\sqrt{\mu ~( 1 - \beta )}}{a_{\beta}^{3/2}} ~ \frac{1}{2 \pi} \int_{0}^{2 \pi}
g(f_{\beta}) ~\frac{r^{2}}{\sqrt{\mu ~( 1 - \beta ) ~p_{\beta}}}~ df_{\beta}
\nonumber \\
&=& \frac{1}{a_{\beta}^{2}~ \sqrt{1 - e_{\beta}^{2}}} ~\frac{1}{2 \pi} ~
  \int_{0}^{2 \pi} ~ g(f_{\beta}) ~r^{2} ~ df_{\beta} ~,
\end{eqnarray}
assuming non-pseudo-circular orbits and the fact that orbital elements exhibit
only small changes during the time interval $T$; $a_{\beta}$ is semi-major axis,
$e_{\beta}$ is eccentricity, $f_{\beta}$ is true anomaly,
$p_{\beta} = a_{\beta}(1 - e_{\beta}^{2})$;
the second and the third Kepler's laws were used:
$r^{2} ~df_{\beta}/dt = \sqrt{\mu ( 1 - \beta ) p_{\beta}}$ --
conservation of angular momentum,
$a_{\beta}^{3}/T^{2} = \mu ( 1 - \beta ) / (4 \pi^{2})$.
(For more details see Kla\v{c}ka, 1992b.)

Rewriting Eq. (2) into the form
\begin{equation}\label{4}
\frac{d\vec{v}}{dt} = - \frac{\mu ~\left ( 1 - \beta \right )}{r^{2}} \vec{S}
		      - \beta \frac{\mu}{r^{2}}
    \left \{
    \left ( \frac{\vec{v} \cdot \vec{S}}{c} \right ) \vec{S}
    + \frac{\vec{v}}{c} \right \} ~,
\end{equation}
we can immediately write for components of perturbation acceleration
to Keplerian motion:
\begin{equation}\label{5}
F_{\beta ~R} = -~2~ \beta ~\frac{\mu}{r^{2}} ~\frac{v_{R}}{c} ~,~~
F_{\beta ~T} = -~ \beta ~\frac{\mu}{r^{2}} ~\frac{v_{T}}{c} ~,~~
F_{\beta ~N} = 0 ~,
\end{equation}
where $F_{\beta ~R}$, $F_{\beta ~T}$ and $F_{\beta ~N}$ are radial, transversal
and normal components of perturbation acceleration, and
two-body problem yields
\begin{eqnarray}\label{6}
v_{R} &=& \sqrt{\frac{\mu ~( 1 - \beta )}{p_{\beta}}} ~ e_{\beta} \sin f_{\beta} ~,~
\nonumber \\
v_{T} &=& \sqrt{\frac{\mu ~( 1 - \beta )}{p_{\beta}}} ~
\left ( 1 + e_{\beta} \cos f_{\beta} \right ) ~.
\end{eqnarray}
The important fact that pertubation acceleration is proportional to
$v/c$ ($\ll$ 1) ensures the above metioned small changes of orbital elements
during the time interval $T$.

Perturbation equations of celestial mechanics yield for osculating orbital
elements ($a_{\beta}$ -- semi-major axis; $e_{\beta}$ -- eccentricity;
$i_{\beta}$ -- inclination
(of the orbital plane to the reference frame);
$\Omega_{\beta}$ -- longitude of the ascending node; $\omega_{\beta}$ --
longitude of pericenter; $\Theta_{\beta}$ --
$\Theta_{\beta} = \omega_{\beta} + f_{\beta}$
is the position angle of the particle on the orbit,
when measured from the ascending node in the direction of the particle's
motion):
\begin{eqnarray}\label{7}
\frac{d a_{\beta}}{d t} &=& \frac{2~a_{\beta}}{1~-~e_{\beta}^{2}} ~
	  \sqrt{\frac{p_{\beta}}{\mu \left ( 1 ~-~ \beta \right )}} ~
	  \left \{
	  F_{\beta ~R} ~e_{\beta}~ \sin f_{\beta} +
	  F_{\beta ~T} \left ( 1~+~e_{\beta}~ \cos f_{\beta} \right ) \right \} ~,
\nonumber \\
\frac{d e_{\beta}}{d t} &=&
	  \sqrt{\frac{p_{\beta}}{\mu \left ( 1 ~-~ \beta \right )}} ~ \left \{
	  F_{\beta ~R} ~ \sin f_{\beta} +
	  F_{\beta ~T} \left [ \cos f_{\beta} ~+~
	 \frac{e_{\beta} +  \cos f_{\beta}}{1 + e_{\beta} \cos f_{\beta}}
		  \right ] \right \} ~,
\nonumber \\
\frac{d i_{\beta}}{d t} &=& \frac{r}{\sqrt{\mu \left ( 1 ~-~ \beta \right ) p_{\beta}}} ~
		    F_{\beta ~N} ~ \cos \Theta_{\beta} ~,
\nonumber \\
\frac{d \Omega_{\beta}}{d t} &=&
	  \frac{r}{\sqrt{\mu \left ( 1 ~-~ \beta \right ) p_{\beta}}} ~
	  F_{\beta ~N} ~ \frac{\sin \Theta_{\beta}}{\sin i_{\beta}} ~,
\nonumber \\
\frac{d \omega_{\beta}}{d t} &=& -~ \frac{1}{e_{\beta}} ~
	  \sqrt{\frac{p_{\beta}}{\mu \left ( 1 ~-~ \beta \right )}} ~ \left \{
	  F_{\beta ~R} \cos f_{\beta} - F_{\beta ~T}
	  \frac{2 + e_{\beta} \cos f_{\beta}}{1 + e_{\beta} \cos f_{\beta}}
	  \sin f_{\beta} \right \} ~-~
\nonumber \\
& &	  \frac{r}{\sqrt{\mu \left ( 1 ~-~ \beta \right ) p_{\beta}}} ~
	  F_{\beta ~N} ~ \frac{\sin \Theta_{\beta}}{\sin i_{\beta}} ~\cos i_{\beta} ~,
\nonumber \\
\frac{d \Theta_{\beta}}{d t} &=&
	  \frac{\sqrt{\mu \left ( 1 ~-~ \beta \right ) p_{\beta}}}{r^{2}} ~-~
	  \frac{r}{\sqrt{\mu \left ( 1 ~-~ \beta \right ) p_{\beta}}} ~
	  F_{\beta ~N} ~ \frac{\sin \Theta_{\beta}}{\sin i_{\beta}} ~\cos i_{\beta} ~,
\end{eqnarray}
where $r = p_{\beta} / (1 + e_{\beta} \cos f_{\beta})$.

Inserting Eqs. (5) -- (6) into Eq. (7), one easily obtains
\begin{eqnarray}\label{8}
\frac{da_{\beta}}{dt} &=& -~\beta \frac{\mu}{r^{2}} \frac{2 a_{\beta}}{c}
    \frac{1 + e_{\beta}^{2} + 2 e_{\beta} \cos f_{\beta}
    + e_{\beta}^{2}  \sin^{2} f_{\beta}}{1~-~e_{\beta}^{2}} ~,
\nonumber \\
\frac{de_{\beta}}{dt} &=& -~ \beta \frac{\mu}{r^{2}} \frac{1}{c}  \left (
      2 e_{\beta} + e_{\beta}  \sin^{2} f_{\beta} + 2 \cos f_{\beta} \right ) ~,
\nonumber \\
\frac{d i_{\beta}}{dt} &=& 0 ~,
\nonumber \\
\frac{d\Omega_{\beta}}{dt} &=& 0 ~,
\nonumber \\
\frac{d\omega_{\beta}}{dt} &=& \beta \frac{\mu}{r^{2}} \frac{1}{c}
    \frac{1}{e_{\beta}} ~ \left (
    2  - e_{\beta} \cos f_{\beta} \right ) \sin f_{\beta} ~,
\nonumber \\
\frac{d \Theta_{\beta}}{dt} &=& \frac{\sqrt{\mu \left ( 1 - \beta \right ) p_{\beta}}}{r^{2}} ~.
\end{eqnarray}
It is worth mentioning that $da_{\beta}/dt <$ 0 for any time $t$ (we again stress
that near circular orbits are not considered).

The set of differential equations Eqs. (8) has to be complemented
with initial conditions. If the subscript $0$ denotes orbital elements
of the parent body (e. g., comet), then even if a particle is ejected with
zero relative velocity, it will move on a different orbit due to the
central force on it being reduced because of the radiation.
(For more general case, with non-zero ejection, see
Gajdo\v{s}\'{\i}k and Kla\v{c}ka 1999.)
The elements of the "new" orbit are given by
\begin{equation}\label{9}
a_{\beta ~in} = a_{0} \left ( 1 - \beta \right ) \left ( 1 - 2 \beta
       \frac{1 + e_{0} \cos f_{0}}{1~-~e_{0}^{2}}
       \right ) ^{-1} ~,
\end{equation}
\begin{equation}\label{10}
e_{\beta ~in} = \sqrt{1 - \frac{1 - e_{0}^{2} - 2 \beta \left (
      1 + e_{0} \cos f_{0} \right )}{
      \left ( 1 - \beta \right )^{2}}} ~,
\end{equation}
where $f_{0} \equiv \Theta_{0} - \omega_{0}$,
$\omega _{\beta ~in}$ has to be obtained from
\begin{eqnarray}\label{11}
e_{\beta~ in} ~ \cos \left ( \Theta _{0} - \omega _{\beta ~in} \right ) &=&
  \frac{\beta + e_{0} \cos f_{0}}{1 - \beta} ~,
\nonumber \\
e_{\beta ~in} ~ \sin \left ( \Theta _{0} - \omega _{\beta ~in} \right ) &=&
  \frac{e_{0} \sin f_{0}}{1 - \beta} ~,
\end{eqnarray}
\begin{equation}\label{12}
\Omega_{\beta~ in} = \Omega_{0} ~, ~~i_{\beta~ in} = i_{0} ~,~~
\Theta_{\beta ~in} = \Theta_{0} ~.
\end{equation}
Some figures may be found in Kla\v{c}ka (1992b).

By inserting Eqs. (8) into Eq. (3), taking into account that $e_{\beta ~in} <$ 1
(see Eq. (10)), one can easily obtain the secular changes of orbital elements:
\begin{equation}\label{13}
\langle  \frac{d a_{\beta}}{d t} \rangle = -~ \beta ~ \frac{\mu}{c} ~
       \frac{2 + 3 e_{\beta}^{2}}{a_{\beta}~
	 \left ( 1 - e_{\beta}^{2} \right )^{3/2}} ~,
\end{equation}
\begin{equation}\label{14}
\langle  \frac{d e_{\beta}}{d t} \rangle = -~ \frac{5}{2}~ \beta ~ \frac{\mu}{c} ~
       \frac{e_{\beta}}{a_{\beta}^{2}~
	 \left ( 1 - e_{\beta}^{2} \right )^{1/2}} ~,
\end{equation}
\begin{equation}\label{15}
\langle  \frac{d \Theta_{\beta}}{d t} \rangle =
	\frac{\sqrt{\mu ~ \left ( 1 - \beta \right )}}{a_{\beta}^{3/2}} ~,
\end{equation}
\begin{equation}\label{16}
\langle  \frac{d i_{\beta}}{d t} \rangle =
\langle  \frac{d \Omega_{\beta}}{d t} \rangle =
\langle  \frac{d \omega_{\beta}}{d t} \rangle = 0 ~.
\end{equation}

\subsection{Advancement of perihelion, $v^{2}/c^{2}$-terms}

Working to first order, as above, in $v/c$ yields no secular change of $\omega$
(advancement of perihelion in Solar System). Thus, we have to include
higher orders. Including now terms up to  $v^{2}/c^{2}$, Eq. (1) yields
\begin{equation}\label{17}
\frac{d \vec{v}}{d t} = \beta ~ \frac{\mu}{r^{2}} \left \{ - ~\frac{1}{2}
   \left ( \frac{\vec{v}}{c} \right ) ^{2} \vec{S} +
   \left ( \frac{\vec{v} \cdot \vec{S}}{c} \right )
   \frac{\vec{v}}{c} \right \} ~,
\end{equation}
which gives
\begin{equation}\label{18}
F_{\beta~ R} = \beta \frac{\mu}{r^{2}} \frac{v_{R}^{2} - v_{T}^{2}}{2 c^{2}} ~,~
F_{\beta~T} = \beta \frac{\mu}{r^{2}} \frac{v_{R} v_{T}}{c^{2}} ~.
\end{equation}
In reality, these components of perturbation acceleration represent only a part
of the total perturbation acceleration, since general relativity theory
produces other terms. The total
advancement of pericenter is given (compare Balek and Kla\v{c}ka 2002) as
\begin{equation}\label{19}
\langle  \frac{d \omega_{\beta}}{d t} \rangle  = \frac{3 \mu ^{3/2}}
       {c^{2} a_{\beta}^{5/2}  \left ( 1 - e_{\beta}^{2} \right )}
	\frac{1 + \beta ^{2} \left ( - 13 / 8 + 7 / e_{\beta} \right ) / 3}{
       \left ( 1 - \beta \right ) ^{1/2}} ~.
\end{equation}

It can be easily verified that $< d \omega_{\beta} ~/~ d t >$ is
i) an increasing function of $\beta$,
ii) the perihelion circulates in a positive direction,
iii) the rate of the advancement of perihelion is not bounded for
$\beta \rightarrow$ 1.

\section{Electromagnetic radiation and general equation of motion}
General equation of motion of a particle under the action of electromagnetic
radiation can be written in the form (Kla\v{c}ka 2002c)
\begin{equation}\label{20}
\frac{d ~p^{\mu}}{d~ \tau} = \sum_{j=1}^{3} \left (
		  \frac{w_{1}^{2} ~S ~A'}{c} ~Q_{j} ' ~+~ Q_{ej} '  \right ) ~
			       \left ( b_{j}^{\mu} ~-~ u^{\mu} ~/~c  \right ) ~,
\end{equation}
where also nonradial components (in the proper frame of reference of the particle)
of radiation pressure are considered and also thermal emission of dust
particle is taken into account ($d m / d \tau =$ 0).

Let us consider orbital evolution of real dust particle,
under the action of gravitational and electromagnetic radiation fields of the
central body (star). We can write for the Solar System (Kla\v{c}ka 2002c)
\begin{eqnarray}\label{21}
\frac{d~ \vec{v}}{d ~t} &=& -~ \frac{G~M_{\odot}}{r^{2}} ~ \vec{e}_{1} ~+~
\nonumber   \\
& &	      \frac{G~M_{\odot}}{r^{2}} ~
	      \sum_{j=1}^{3} ~\beta_{j} ~\left [  \left ( 1~-~ 2~
	      \frac{\vec{v} \cdot \vec{e}_{1}}{c} ~+~
	      \frac{\vec{v} \cdot \vec{e}_{j}}{c} \right ) ~ \vec{e}_{j}
	      ~-~ \frac{\vec{v}}{c} \right ] ~+~
\nonumber   \\
& &	      \sum_{j=1}^{3} ~Q'_{ej} ~\left [  \left ( 1~+~
	      \frac{\vec{v} \cdot \vec{e}_{j}}{c} \right ) ~ \vec{e}_{j}
	      ~-~ \frac{\vec{v}}{c} \right ]  ~,
\end{eqnarray}
where
\begin{eqnarray}\label{22}
\vec{e}_{j} &=& ( 1 ~-~ \vec{v} \cdot \vec{e}'_{j} / c ) ~ \vec{e}'_{j}  ~+~
	      \vec{v} / c ~, ~~j = 1, 2, 3 ~,
\nonumber   \\
\beta_{1} &=& \frac{\pi ~R_{\odot}^{2}}{G~M_{\odot}~m~c}
	      \int_{0}^{\infty} B_{\odot} ( \lambda ) \left \{
	      C'_{ext} ( \lambda / w ) - C'_{sca} ( \lambda / w ) ~
	      g'_{1} ( \lambda / w ) \right \}  d \lambda ~,
\nonumber   \\
\beta_{2} &=& \frac{\pi ~R_{\odot}^{2}}{G~M_{\odot}~m~c}
	      \int_{0}^{\infty} B_{\odot} ( \lambda ) \left \{
	      ~-~ C'_{sca} ( \lambda / w ) ~
	      g'_{2} ( \lambda / w ) \right \}  d \lambda ~,
\nonumber   \\
\beta_{3} &=& \frac{\pi ~R_{\odot}^{2}}{G~M_{\odot}~m~c}
	      \int_{0}^{\infty} B_{\odot} ( \lambda ) \left \{
	      ~-~ C'_{sca} ( \lambda / w ) ~
	      g'_{3} ( \lambda / w ) \right \}  d \lambda ~,
\nonumber   \\
w &=& 1~-~\vec{v} \cdot \vec{e}_{1} ~/~ c ~,
\end{eqnarray}
$R_{\odot}$ denotes the radius of the Sun and $B_{\odot} ( \lambda )$ is
the solar radiance at a wavelength of $\lambda$; $G$, $M_{\odot}$, and
$r$ are the gravitational constant, the mass of the Sun, and the distance
of the particle from the center of the Sun, respectively.
The asymmetry parameter vector $\vec{g}'$ is defined by $\vec{g}'$ $=$
$( 1 / C'_{sca} ) \int \vec{n}' ( d C'_{sca} / d \Omega ') d \Omega '$, where
$\vec{n}'$ is a unit vector in the direction of scattering;
$\vec{g}'$ $=$ $g'_{1} ~ \vec{e}'_{1}$ $+$ $g'_{2} ~ \vec{e}'_{2}$ $+$
$g'_{3} ~ \vec{e}'_{3}$, $\vec{e}'_{1}$ $=$
$( 1 ~+~ \vec{v} \cdot \vec{e}_{1} / c ) ~ \vec{e}_{1}$  $-$
$\vec{v} / c$, $\vec{e}'_{i} \cdot \vec{e}'_{j} = \delta _{ij}$.

The first set of Eqs. (22) is important from the point of view that general
equation of motion represented by Eqs. (20) and (21) has to be reduced to
the special case treated by Einstein (1905).

We see that the motion of real, arbitrarily shaped, particle may by much more
complicated than that corresponding to the P-R effect; P-R effect is a special
case of Eq. (20). To the first order in $v/c$, Eq. (21) reduces to Eq. (2)
for $\beta_{2}$ $=$ $\beta_{3}$ $=$ $Q'_{e1}$ $=$ $Q'_{e2}$
$=$ $Q'_{e3}$ $=$ 0; $\beta_{1} \equiv \beta$, when comparing Eq. (2) and
Eq. (21). Motion of a real particle depends on optical properties of the
particle, e. g., shape, chemical composition of the particle, mass
distribution within the particle (porosity).

\section{P-R effect and secular changes of orbital elements --
gravitation as a central acceleration}
If one would like to inspire with calculation od orbital elements
as it was presented in section 3, he should consider
$-~\mu ~( 1~-~\beta_{1} )~\vec{S}~/~r^{2}$
as a central acceleration determining
osculating orbital elements even for general case represented by Eq. (21).
However, the quantity $\beta_{1}$ depends on optical properties of the particle
and, thus, also on particle's position with respect to the source of
electromagnetic radiation.

It is not wise to use $-~\mu ~( 1~-~\beta_{1} )~\vec{S}~/~r^{2}$
as a central acceleration determining
osculating orbital elements for general case represented by Eq. (21), since
$\mu ~( 1~-~\beta_{1} )$ changes almost randomly during a motion. Thus, it is
not wise to use $-~\mu ~( 1~-~\beta )~\vec{S}~/~r^{2}$
as a central acceleration determining
osculating orbital elements for the P-R effect, if we want to compare
the evolution of orbital elements for the P-R effect and the evolution of
orbital elements for general case represented by Eq. (21). We need something
which is not changing almost randomly. The wise quantity is $\mu$. Thus,
central acceleration
$-~\mu ~\vec{S}~/~r^{2}$ will be used as a central acceleration
determining osculating orbital elements.

\subsection{P-R effect and perturbation oquations of celestial mechanics}
We use $-~\mu ~\vec{S}~/~r^{2}$ as a central
acceleration determining osculating orbital elements.

Taking into account Eq. (2), we take the action of electromagnetic radiation
as a pertubation to the two-body problem.
We can immediately write for components of perturbation acceleration:
\begin{equation}\label{23}
F_{R} = \beta ~\frac{\mu}{r^{2}} ~-
~2~ \beta ~\frac{\mu}{r^{2}} ~\frac{v_{R}}{c} ~,~~
F_{T} = -~ \beta ~\frac{\mu}{r^{2}} ~\frac{v_{T}}{c} ~,~~
F_{N} = 0 ~,
\end{equation}
where $F_{R}$, $F_{T}$ and $F_{N}$ are radial, transversal
and normal components of perturbation acceleration, and
two-body problem yields
\begin{eqnarray}\label{24}
v_{R} &=& \sqrt{\frac{\mu}{p}} ~ e \sin f ~,~
\nonumber \\
v_{T} &=& \sqrt{\frac{\mu}{p}} ~ \left ( 1 + e \cos f \right ) ~.
\end{eqnarray}

Perturbation equations of celestial mechanics yield for osculating orbital
elements ($a$ -- semi-major axis; $e$ -- eccentricity; $i$ -- inclination
(of the orbital plane to the reference frame);
$\Omega$ -- longitude of the ascending node; $\omega$ --
longitude of pericenter; $\Theta$ --
$\Theta = \omega + f$ is the position angle of the particle on the orbit,
when measured from the ascending node in the direction of the particle's
motion):
\begin{eqnarray}\label{25}
\frac{d a}{d t} &=& \frac{2~a}{1~-~e^{2}} ~
	  \sqrt{\frac{p}{\mu}} ~ \left \{
	  F_{R} ~e~ \sin f +
	  F_{T} \left ( 1~+~e~ \cos f \right ) \right \} ~,
\nonumber \\
\frac{d e}{d t} &=& \sqrt{\frac{p}{\mu}} ~ \left \{ F_{R} ~ \sin f +
		    F_{T} \left [ \cos f ~+~ \frac{e +	\cos f}{1 + e \cos f}
		    \right ] \right \} ~,
\nonumber \\
\frac{d i}{d t} &=& \frac{r}{\sqrt{\mu ~p}} ~
		    F_{N} ~ \cos \Theta ~,
\nonumber \\
\frac{d \Omega}{d t} &=& \frac{r}{\sqrt{\mu ~p}} ~
			 F_{N} ~ \frac{\sin \Theta}{\sin i} ~,
\nonumber \\
\frac{d \omega}{d t} &=& -~ \frac{1}{e} ~\sqrt{\frac{p}{\mu}} ~ \left \{
			 F_{R} \cos f - F_{T} \frac{2 + e \cos f}{1 + e \cos f}
			 \sin f \right \} ~-~
\nonumber \\
& &			 \frac{r}{\sqrt{\mu ~p}} ~
			 F_{N} ~ \frac{\sin \Theta}{\sin i} ~\cos i ~,
\nonumber \\
\frac{d \Theta}{d t} &=& \frac{\sqrt{\mu ~p}}{r^{2}} ~-~
			 \frac{r}{\sqrt{\mu ~p}} ~
			 F_{N} ~ \frac{\sin \Theta}{\sin i} ~\cos i ~,
\end{eqnarray}
where $r = p / (1 + e \cos f)$.

Inserting Eqs. (23) -- (24) into Eq. (25), one easily obtains
\begin{eqnarray}\label{26}
\frac{da}{dt} &=& \frac{2~ \beta}{r^{2}} ~\sqrt{\frac{\mu~ a^{3}}{1~-~e^{2}}} ~
    e ~\sin f ~-~
    \beta \frac{\mu}{r^{2}} ~\frac{2~a}{c}~
    \frac{1 + e^{2} + 2 e \cos f + e^{2}  \sin^{2} f}{1~-~e^{2}} ~,
\nonumber \\
\frac{de}{dt} &=& \beta ~ \frac{\sqrt{\mu~p}}{r^{2}} ~ \sin f ~-~
      \beta ~ \frac{\mu}{r^{2}} \frac{1}{c}  \left (
      2 e + e  \sin^{2} f + 2 \cos f \right ) ~,
\nonumber \\
\frac{d i}{dt} &=& 0 ~,
\nonumber \\
\frac{d\Omega}{dt} &=& 0 ~,
\nonumber \\
\frac{d\omega}{dt} &=& -~ \beta ~ \frac{\sqrt{\mu~p}}{r^{2}} ~\frac{1}{e} ~
		       \cos f ~-~ \beta ~ \frac{\mu}{r^{2}} ~ \frac{1}{c}
		       \frac{1}{e} ~ \left ( 2	- e \cos f \right ) ~ \sin f ~,
\nonumber \\
\frac{d \Theta}{dt} &=& \frac{\sqrt{\mu ~p}}{r^{2}} ~.
\end{eqnarray}
It is worth mentioning that $da/dt <$ 0 for any time $t$ does not hold
(perturbation corresponds to complete nongravitational acceleration, and, thus,
Eqs. (8) do not hold).

The set of differential equations Eqs. (26) has to be complemented
with initial conditions. If the subscript $0$ denotes orbital elements
of the parent body (e. g., comet), then particle ejected with zero relative
velocity will have initial osculating orbital elements identical with those
of the parent body:
\begin{eqnarray}\label{27}
a_{in} &=& a_{0} ~, ~~e_{in} = e_{0} ~,~~i_{in} = i_{0} ~,
\nonumber \\
\Omega_{in} &=& \Omega_{0} ~,~~\omega_{in} = \omega_{0} ~,~~
\Theta_{in} = \Theta_{0} ~.
\end{eqnarray}

The important fact is that Eqs. (26) contain also terms not proportional
to $v/c$ ($\ll$ 1). These important terms protect us to use procedure
analogous to that represented by Eq. (3). While dispersion of osculating orbital
elements is very small during a time interval $T$ for the case when
$\mu ~ ( 1 - \beta )$ is used in central acceleration, the dispersion of
osculating orbital elements may be large during the same time interval
for the case when $\mu$ is used in central acceleration. Thus, any formal
averaging of Eqs. (26) leading to equations analogous to Eqs. (13)-(16)
is not correct. This fact was discussed in Kla\v{c}ka (1992b),
see also Figs. 1 and 2 in Kla\v{c}ka (1994).

\subsection{Expressions for secular changes of semi-major axis
and eccentricity}
We have already explained that it is not allowed to make a simple time averaging
analogous to that described by Eq. (3), when $-~\mu ~\vec{S}~/~r^{2}$ is used
as a central acceleration determining osculating orbital elements. However, it
is of interest if we have to numerically solve Newtonian vectorial equation of
motion (Eq. (2)) and make numerical time averaging (over a time interval between
passages through two following pericenters), or some analytical simplifications
can be done -- something analogous to Eqs. (13)-(14).

Correct answer is: yes, we can make some analytical simplifications
when we want to obtain secular changes of semi-major axis and eccentricity even
when $-~\mu ~\vec{S}~/~r^{2}$ is used
as a central acceleration determining osculating orbital elements. We will
derive the correct equations in the following two subsections.

\subsubsection{Radial forces and orbital elements}
We will proceed according to Kla\v{c}ka (1994), in this subsection.

Let us consider a gravitational system of two bodies
\begin{equation}\label{28}
\dot{\vec{v}} = -~\frac{\mu}{r^{2}} ~ \vec{e}_{R} ~,
\end{equation}
where $\vec{e}_{R} \equiv \vec{e}_{1} \equiv \vec{S}$. Let perturbation
acceleration exists in the form
\begin{equation}\label{29}
\vec{a} = \beta ~ \frac{\mu}{r^{2}} ~ \vec{e}_{R} ~,
\end{equation}
$0 \le \beta <$ 1. Thus, the final equation of motion is
\begin{equation}\label{30}
\dot{\vec{v}} = -~\frac{\mu ~ \left ( 1 ~-~ \beta \right )}{r^{2}} ~ \vec{e}_{R} ~.
\end{equation}
Eq. (30) yields as a solution the well-known Keplerian motion and the orbit
is given by
\begin{equation}\label{31}
r = \frac{p_{c}}{1 ~+~ e_{c} ~\cos \left ( \Theta ~-~ \omega_{c} \right )} ~,
\end{equation}
where
\begin{equation}\label{32}
p_{c} = a_{c} ~ \left ( 1 ~-~ e_{c}^{2} \right ) ~.
\end{equation}
The subscript ``c'' denotes that orbital elements are constants of motion.
If we write
\begin{equation}\label{33}
\vec{v} = v_{R} ~ \vec{e}_{R} ~+~ v_{T} ~ \vec{e}_{T} ~,
\end{equation}
where $\vec{e}_{T}$ is a unit vector transverse to the radial vector
$\vec{e}_{R}$ in the plane of the trajectory (positive in the direction of motion),
we have
\begin{equation}\label{34}
v_{R} = \sqrt{\mu ~ \left ( 1 ~-~ \beta \right )~p_{c}^{-1}} ~e_{c} ~
	\sin \left ( \Theta ~-~ \omega_{c} \right ) ~,
\end{equation}
\begin{equation}\label{35}
v_{T} = \sqrt{\mu ~ \left ( 1 ~-~ \beta \right )~p_{c}^{-1}} ~\left [
	1 ~+~ e_{c} ~
	\cos \left ( \Theta ~-~ \omega_{c} \right ) \right ] ~.
\end{equation}

In principle, we may consider also a new set of orbital elements, which
are defined by the central gravitational acceleration. Equations (31), (32),
(34) and (35) are then of the form
\begin{equation}\label{36}
r = \frac{p}{1 ~+~ e ~\cos \left ( \Theta ~-~ \omega \right )} ~,
\end{equation}
\begin{equation}\label{37}
p = a ~ \left ( 1 ~-~ e^{2} \right ) ~,
\end{equation}
\begin{equation}\label{38}
v_{R} = \sqrt{\mu ~p^{-1}} ~e ~ \sin \left ( \Theta ~-~ \omega \right ) ~,
\end{equation}
\begin{equation}\label{39}
v_{T} = \sqrt{\mu ~p^{-1}} ~\left [ 1 ~+~ e ~
	\cos \left ( \Theta ~-~ \omega \right ) \right ] ~;
\end{equation}
the fact that $\Theta$ is unchanged in both sets of orbital elements is used.

Position vector and velocity vector define a state of the body at any time.
Equations (31) and (36) yield then
\begin{equation}\label{40}
\frac{p_{c}}{1 ~+~ e_{c} ~\cos \left ( \Theta ~-~ \omega_{c} \right )}
= \frac{p}{1 ~+~ e ~\cos \left ( \Theta ~-~ \omega \right )} ~.
\end{equation}
Analogously, the other two pairs of equations (Eqs. (34) and (38), and,
Eqs. (35) and (39)) give
\begin{equation}\label{41}
\sqrt{\left ( 1 ~-~ \beta \right )~p_{c}^{-1}} ~e_{c} ~
	\sin \left ( \Theta ~-~ \omega_{c} \right )
= \sqrt{p^{-1}} ~e ~ \sin \left ( \Theta ~-~ \omega \right ) ~,
\end{equation}
\begin{equation}\label{42}
\sqrt{\left ( 1 ~-~ \beta \right )~p_{c}^{-1}} ~\left [ 1 ~+~ e_{c} ~
	\cos \left ( \Theta ~-~ \omega_{c} \right ) \right ]
= \sqrt{p^{-1}} ~\left [ 1 ~+~ e ~
	\cos \left ( \Theta ~-~ \omega \right ) \right ] ~.
\end{equation}
One can easily obtain, using Eqs. (40) and (42),
\begin{equation}\label{43}
p_{c} ~ \left ( 1 ~-~ \beta \right ) = p ~,
\end{equation}
and Eqs. (41)-(42) yield then
\begin{equation}\label{44}
\left ( 1 ~-~ \beta \right ) ~e_{c} ~
	\sin \left ( \Theta ~-~ \omega_{c} \right ) =
	e~ \sin \left ( \Theta ~-~ \omega \right ) ~,
\end{equation}
\begin{equation}\label{45}
\left ( 1 ~-~ \beta \right ) ~\left [
	1 ~+~ e_{c} ~
	\cos \left ( \Theta ~-~ \omega_{c} \right ) \right ] =
	1 ~+~ e ~ \cos \left ( \Theta ~-~ \omega \right ) ~.
\end{equation}
Eq. (45) can be written in the form
\begin{equation}\label{46}
\left ( 1 ~-~ \beta \right ) ~ e_{c} ~
	\cos \left ( \Theta ~-~ \omega_{c} \right ) ~-~ \beta  =
	e~ \cos \left ( \Theta ~-~ \omega \right ) ~.
\end{equation}
Eqs. (44) and (46) yield
\begin{equation}\label{47}
e^{2} = \left ( 1 ~-~ \beta \right ) ^{2} ~ e_{c}^{2} ~+~ \beta ^{2} ~-~
	2~ \beta ~ \left ( 1 ~-~ \beta \right ) ~ e_{c} ~
	\cos \left ( \Theta ~-~ \omega_{c} \right ) ~.
\end{equation}
Equation for $\omega$ is given by Eqs. (44) and (46), using also Eq. (47).
Finally, Eqs. (32), (37), (43) and (47) yield
\begin{equation}\label{48}
a = a_{c} ~ \left \{ 1 ~+~ \beta ~ \frac{1 ~+~ e_{c}^{2} ~+~ 2~e_{c} ~
	\cos \left ( \Theta ~-~ \omega_{c} \right )}{1~-~e_{c}^{2}} \right \}
	^{-1}  ~.
\end{equation}
Eqs. (47)-(48) show that orbital osculating elments $a$ and $e$ change in time,
during an orbital revolution -- the larger $\beta$, the larger change of
$a$ and $e$.

The osculating orbital elements $e$ and $a$ obtain values between their
maxima and minima, which can be easily found from Eqs. (47)-(48). One can
easily verify that these relations hold:
\begin{equation}\label{49}
e_{min} = | \left ( 1 ~-~ \beta \right ) ~e_{c} ~-~ \beta | \le e \le
	  \left ( 1 ~-~ \beta \right ) ~e_{c} ~+~ \beta =
	  e_{max} ~,
\end{equation}
\begin{equation}\label{50}
\frac{a_{min}}{a_{c}} =
     \frac{1~-~e_{c}}{1~-~e_{c} ~+~ \beta ~\left ( 1 ~+~ e_{c} \right )}
\le \frac{a}{a_{c}} \le
     \frac{1~+~e_{c}}{1~+~e_{c} ~+~ \beta ~\left ( 1 ~-~ e_{c} \right )}
= \frac{a_{max}}{a_{c}} ~.
\end{equation}

\subsubsection{Mean values of semi-major axis and eccentricity}
Eqs. (49) and (50) represent interval of values for eccentricity and semi-major
axis, when $-~\mu ~\vec{S}~/~r^{2}$ is used
as a central acceleration determining osculating orbital elements. However,
we can make time averaging during a period $T$, which was decribed by Eq. (3):
\begin{eqnarray}\label{51}
\langle e \rangle &=& \frac{1}{T} ~ \int_{0}^{T} e(t) ~dt ~, ~~
\langle a \rangle = \frac{1}{T} ~ \int_{0}^{T} a(t) ~dt ~,
\nonumber \\
r^{2} ~ \frac{d f_{c}}{d t} &=&
\sqrt{\mu ~\left ( 1 ~-~ \beta \right )~p_{c}} ~, ~~
\frac{a_{c}^{3}}{T^{2}} = \frac{\mu ~ \left ( 1 ~-~ \beta \right )}{4~\pi^{2}} ~, ~~
r = \frac{p_{c}}{1 ~+~ e_{c} ~\cos f_{c}} ~.
\end{eqnarray}
Eqs. (51) yield
\begin{eqnarray}\label{52}
\langle e \rangle &=& \left ( 1 ~-~ e_{c}^{2} \right ) ^{3/2} ~
	 \frac{1}{2~\pi} ~ \int_{0}^{2 \pi}
\frac{e \left ( f_{c} \right )}{ \left ( 1 ~+~ e_{c} ~\cos f_{c} \right )^{2}}
~df_{c} ~,
\nonumber \\
\langle a \rangle &=& \left ( 1 ~-~ e_{c}^{2} \right ) ^{3/2} ~
	 \frac{1}{2~\pi} ~ \int_{0}^{2 \pi}
\frac{a \left ( f_{c} \right )}{ \left ( 1 ~+~ e_{c} ~\cos f_{c} \right )^{2}}
~df_{c} ~.
\end{eqnarray}
Using Eqs. (47) and (48), we finally obtain
\begin{eqnarray}\label{53}
\langle e \rangle &=& \left ( 1 - e_{c}^{2} \right ) ^{3/2} ~
	 \frac{1}{2 \pi} ~ \int_{0}^{2 \pi}
\frac{\sqrt{\left ( 1 - \beta \right ) ^{2}  e_{c}^{2} + \beta ^{2} -
	2 \beta \left ( 1 - \beta \right )  e_{c}
	\cos f_{c}}}{\left ( 1 + e_{c} \cos f_{c} \right )^{2}}
	~df_{c} ~,
\nonumber \\
\langle a \rangle &=& a_{c}  \left ( 1 - e_{c}^{2} \right ) ^{3/2} ~
	 \frac{1}{2 \pi} ~ \int_{0}^{2 \pi}
	\frac{\left [ 1 + \beta \left ( 1 + e_{c}^{2} + 2 e_{c}
	\cos f_{c} \right ) / \left ( 1 - e_{c}^{2} \right ) \right ]^{-1}}
	{\left ( 1 + e_{c} \cos f_{c} \right )^{2}}
	~df_{c} ~.
\end{eqnarray}

The following properties can be verified: \\
i) $\langle e \rangle \ge \beta$,
$\langle e \rangle = \beta$ for $e_{c} =$ 0; ~
ii) $\langle e \rangle / e_{c} \ge$ 1,
$\langle e \rangle = e_{c}$ for $\beta =$ 0; \\
iii) $\partial \langle e \rangle / \partial e_{c} >$ 0;~
iv) $\partial \langle e \rangle / \partial \beta >$ 0; \\
v) $\langle a \rangle \ge a_{c} / ( 1 + \beta )$,
$\langle a \rangle = a_{c} / ( 1 + \beta )$ for $e_{c} =$ 0; \\
vi) $\langle a \rangle / a_{c} \le$ 1,
$\langle a \rangle = a_{c}$ for $\beta =$ 0; \\
vii) $\partial \langle a \rangle / \partial e_{c} >$ 0;~
viii) $\partial \langle a \rangle / \partial \beta <$ 0;~
ix) $\partial \langle a \rangle / \partial a_{c} >$ 0.

\subsubsection{Secular changes of semi-major axis and eccentricity}
Summarizing our results, it is possible to calculate secular evolution of
eccentricity and semi-major axis, according to the following prescription.

At first, initial conditions for $a_{\beta}$ and $e_{\beta}$ are calculated:
\begin{eqnarray}\label{54}
\left ( a_{\beta} \right )_{in} &=& a_{0} \left ( 1 - \beta \right )
       \left ( 1 - 2 \beta \frac{1 + e_{0} \cos f_{0}}{1~-~e_{0}^{2}}
       \right ) ^{-1} ~,
\nonumber \\
\left ( e_{\beta} \right )_{in} &=& \sqrt{1 - \frac{1 - e_{0}^{2} - 2 \beta \left (
      1 + e_{0} \cos f_{0} \right )}{\left ( 1 - \beta \right )^{2}}} ~,
\end{eqnarray}
supposing that particle was ejected with zero ejection velocity
from a parent body -- quantities
with subscript $0$ belongs to the parent body's trajectory.

As the second step, the set of the following differential equations must
be solved for the above presented initial conditions:
\begin{eqnarray}\label{55}
\frac{d a_{\beta}}{d t} &=& -~ \beta ~ \frac{\mu}{c} ~
       \frac{2 + 3 e_{\beta}^{2}}{a_{\beta}~ \left ( 1 - e_{\beta}^{2} \right )^{3/2}} ~,
\nonumber \\
\frac{d e_{\beta}}{d t} &=& -~ \frac{5}{2}~ \beta ~ \frac{\mu}{c} ~
       \frac{e_{\beta}}{a_{\beta}^{2}~ \left ( 1 - e_{\beta}^{2} \right )^{1/2}} ~.
\end{eqnarray}

Finally, semi-major axis and eccentricity are calculated from:
\begin{eqnarray}\label{56}
a &=& a_{\beta} \left ( 1 - e_{\beta}^{2} \right ) ^{3/2} ~
	 \frac{1}{2 \pi} ~ \int_{0}^{2 \pi}
	\frac{\left [ 1 + \beta \left ( 1 + e_{\beta}^{2} + 2 e_{\beta}
	\cos x \right ) / \left ( 1 - e_{\beta}^{2} \right ) \right ]^{-1}}
	{\left ( 1 + e_{\beta} \cos x \right )^{2}}
	~dx ~,
\nonumber \\
e &=& \left ( 1 - e_{\beta}^{2} \right ) ^{3/2} ~
	 \frac{1}{2 \pi} ~ \int_{0}^{2 \pi}
\frac{\sqrt{\left ( 1 - \beta \right ) ^{2}  e_{\beta}^{2} + \beta ^{2} -
	2 \beta \left ( 1 - \beta \right )  e_{\beta}
	\cos x}}{\left ( 1 + e_{\beta} \cos x \right )^{2}}
	~dx ~.
\end{eqnarray}

The set of equations represented by Eqs. (54)-(56) fully corresponds to
detailed numerical calculations of vectorial equation of motion, if
we are interested in secular evolution of eccentricity and semi-major
axis (supposing $(e_{\beta})_{in} <$ 1 and $e_{\beta}$ does not correspond to
pseudo-circular orbit)
for the case when central acceleration is defined by gravity alone.

It is worth mentioning that instantaneous time derivatives of semi-major axis
and eccentricity may be both positive and negative (see Eqs. (26)),
while secular evolution yields that semi-major axis and
eccentricity are decreasing functions of time.

\section{Discussion}
When the orbits are near circular (pseudo-circular)
-- central acceleration contains radiation pressure term -- the orbit can not
reduce in semi-major axis without increasing in eccentricity.
Both types of orbital elements, defined by central accelerations,
have been used in detail numerical calculations in papers
by Kla\v{c}ka and Kaufmannov\'{a} (1992, 1993). Due to the property
$\langle e \rangle \ge \beta$ (see section 5.2.2), one must be aware
that even value $\langle e \rangle > \beta$ may correspond to pseudo-circular
orbit. The results for pseudo-circular orbits were analytically confirmed by
Breiter and Jackson (1998).
Although the analytical approach reproduces known results without detail
numerical calculations, it paralelly produces nonphysical results which may not
be distinguishable from the correct results. The nonphysical analytical results
are caused by usage of P-R effect in the first order in $v/c$ -- very special
analytical solutions will diminish when higher orders in $v/c$ are used.
The nonphysical analytical results have been discussed in more detail by
Kla\v{c}ka (2000a, 2001a).

\section{Conclusion}
We have discussed orbital evolution for the P-R effect.
General equation of motion for interaction between particle and incident
electromagnetic radiation shows that radiation cannot be considered as a part
of central acceleration. The central acceleration has to contain only gravity
of the central body (star). In order to compare secular changes of semi-major
axis and eccentricity for real particle and the P-R effect, the paper presents
secular changes of these orbital elements for the P-R effect (Eqs. (54)-(56)).

\acknowledgements{This paper was supported by the Scientific Grant Agency
VEGA grant No. 1/7067/20.}

\end{document}